\documentclass[draft]{agujournal2019}
\usepackage{apacite}
\usepackage{url} 
\usepackage[italic]{hepnames}
\DeclareGraphicsExtensions{.eps,.png,.jpg,.pdf} 
\usepackage{float}
\usepackage{textcomp} 
\usepackage{lscape} 
\usepackage{multirow} 
\usepackage{mathabx} 
\usepackage{siunitx} 
\usepackage{colortbl} 
\graphicspath{{Figures/}} 
\usepackage{rotating}

\draftfalse
\journalname{arXiv}

\begin{document}

\title{Reference Models for Lithospheric Geoneutrino Signal}

\authors{S. A. Wipperfurth\affil{1}, O. {\v S}r{\'a}mek\affil{2}, W. F. McDonough\affil{1,3}.}

\affiliation{1}{Department of Geology, University of Maryland, College Park, MD 20742, USA}
\affiliation{2}{Department of Geophysics, Faculty of Mathematics and Physics, Charles University, Prague, Czech Republic}
\affiliation{3}{Department of Earth Sciences and Research Center for Neutrino Science, Tohoku University, Sendai 980-8578, Japan}

\correspondingauthor{S. A. Wipperfurth}{swipp@terpmail.umd.edu}

\begin{keypoints}

\item Updated geoneutrino emission model compares results using CRUST2.0, CRUST1.0, LITHO1.0 geophysical models
\item Observed differences between geoneutrino signals from different geophysical models is negligible
\item Necessary improvements to geoneutrino modeling are identified

\end{keypoints}

\begin{abstract}
	
Debate continues on the amount and distribution of radioactive heat producing elements (i.e., U, Th, and K) in the Earth, with estimates for mantle heat production varying by an order of magnitude. Constraints on the bulk-silicate Earth's (BSE) radiogenic power also places constraints on overall BSE composition. Geoneutrino detection is a direct measure of the Earth's decay rate of Th and U. The geoneutrino signal has contributions from the local ($\sim$40\%) and global ($\sim$35\%) continental lithosphere and the underlying inaccessible mantle ($\sim$25\%). Geophysical models are combined with geochemical datasets to predict the geoneutrino signal at current and future geoneutrino detectors. We propagated uncertainties, both chemical and physical, through Monte Carlo methods. Estimated total signal uncertainties are on the order of $\sim$20\%, proportionally with geophysical and geochemical inputs contributing $\sim$30\% and $\sim$70\%, respectively. We find that estimated signals, calculated using CRUST2.0, CRUST1.0, and LITHO1.0, are within physical uncertainty of each other, suggesting that the choice of underlying geophysical model will not change results significantly, but will shift the central value by up to $\sim$15\%, depending on the crustal model and detector location. Similarly, we see no significant difference between calculated layer abundances and bulk-crustal heat production when using these geophysical models. The bulk crustal heat production is calculated as $7 \pm2$~terrawatts, which includes an increase of 1~TW in uncertainty relative to previous studies. Future improvements, including uncertainty attribution and near-field modeling, are discussed. 

\end{abstract}

\section{Introduction}

Estimated at 47 $\pm$ 2 terrawatts (TW) \cite{davies2010}, the deep Earth's radiant heat is primarily comprised of two sources: primordial heat remaining from planetary assembly and core formation, and radiogenic heat produced during nuclear decay of the heat-producing elements (HPE: uranium (U), thorium (Th), and potassium (K)). The radioactive isotopes of these elements --- $^{238}$U (99.3\% g/g of U), $^{235}$U (0.7\%), $^{232}$Th (100\% g/g of Th), $^{40}$K (0.012\% g/g of K) --- presently account for 99\% of Earth's radiogenic heat due to their long half-lives and high abundances relative to other radiogenic elements. 

There are three categories of models which predict the abundance of the HPE's in the bulk-silicate Earth (BSE; crust + mantle ($\sim$0.5\% and $\sim$67\% of Earth by mass, respectively)) and therefore the radiogenic heat production within the BSE. Models which predict low heat production (H) from radiogenic decay ($\approx10$\,TW) are derived from observations of isotopic similarities between Earth and enstatite chondrite \cite{javoy1999,javoy2010} or from models of early Earth collisional erosion of an HPE-enriched crust \cite{oneill2008}. Medium-H models ($\approx20$\,TW) are derived from combining observations from chondrites and mantle melting trends of terrestrial samples \cite{mcdonough1995,palme2014}. Finally, High-H models ($\approx30$\,TW) are derived from simple parameterized mantle convection models \cite{turcotte2014}. There is inconclusive data to evaluate critically the veracity of each of these three BSE models, therefore there is currently not a precise understanding of the composition and thermal evolution of our planet. The Earth's stable isotopic composition is most similar to enstatite chondrites (low-H model), and yet it falls outside of the chondritic defined endmembers in a Fe-Mg-Si plot, indicating Earth is not comprised of a single type of chondrite nor a two component mixture \cite{mcdonough2017}. Furthermore, the use of terrestrial samples and observed conservation of chondritic ratios in terrestrial samples yields a BSE composition (medium-H model) with refractory element abundances (including U and Th, but not K) double that estimated from enstatite chondrites alone. Finally, neither the low-H nor medium-H BSE models satisfy some simple parameterized convection models of the Earth which require larger amounts of radiogenic power to avoid a totally molten mantle for a significant amount of Earth history \cite<high-H model;>{davies1980,schubert1980}. Overall, the range of heat production in BSE models differ by a factor of three (10 to 30 TW). With the consideration of uncertainties and the removal of the HPE contribution from the accessible and HPE enriched continental crust \cite<$7\pm1$~TW;>{huang2013}, these models differ by a factor of thirty in estimates of the radiogenic power in the modern mantle. 

For more than a decade, particle physicists have detected and reported on the Earth's flux of geoneutrinos --- electron antineutrinos ($\bar{\nu_e}$) of terrestrial origin produced during $\beta^-$ decays \cite<$n \rightarrow p^+ + e^- + \bar{\nu_e}$;>[]{araki2005}. The intensity of the geoneutrino flux is proportional to the concentration and sensitive to the spatial distribution of HPEs inside the Earth relative to the detector's location. These elusive particles are exceedingly difficult to detect as they are charge-less leptons with small interaction cross sections of the weak interaction \cite<$\sim10^{-44}~cm^2$;>[]{vogel1999}. Measurement of the geoneutrino flux requires large, underground, scintillation detectors which use the inverse beta decay (IBD) detection process ($\bar{\nu_e} + p^+ \rightarrow n + e^+$). The IBD reaction creates two flashes of light, separated in time ($\sim$200 $\mu$s) and space ($\sim$30 cm), which uniquely classifies the event and provides an energy tag identifying the specific geoneutrino source isotope. The IBD reaction requires an $\bar{\nu_e}$ kinematic threshold energy of 1.806 MeV due to the larger mass of the products relative to the reactants in the production of $\bar{\nu_e}$'s. Consequently, only $\bar{\nu_e}$'s emitted by the decay of U and Th are detectable \cite{araki2005}. The detectors also register other electron antineutrinos present in the geoneutrino energy range (1.8--3.3\,MeV), including reactor antineutrinos. Detectors are located at 1--2~km deep in the upper crust to shield from cosmogenic muons --- the primary non-antineutrino background source. Geoneutrino signals are often reported in terrestrial neutrino units (TNU), with one TNU equal to one IBD detection per 10$^{32}$ free protons ($\approx$ one kiloton of liquid scintillator) in one year with 100\% detection efficiency of a geoneutrino detector. This unit accounts for differences in detector size and efficiency. 

\begin{figure}
    \centering
    \includegraphics[width=1.0\linewidth]{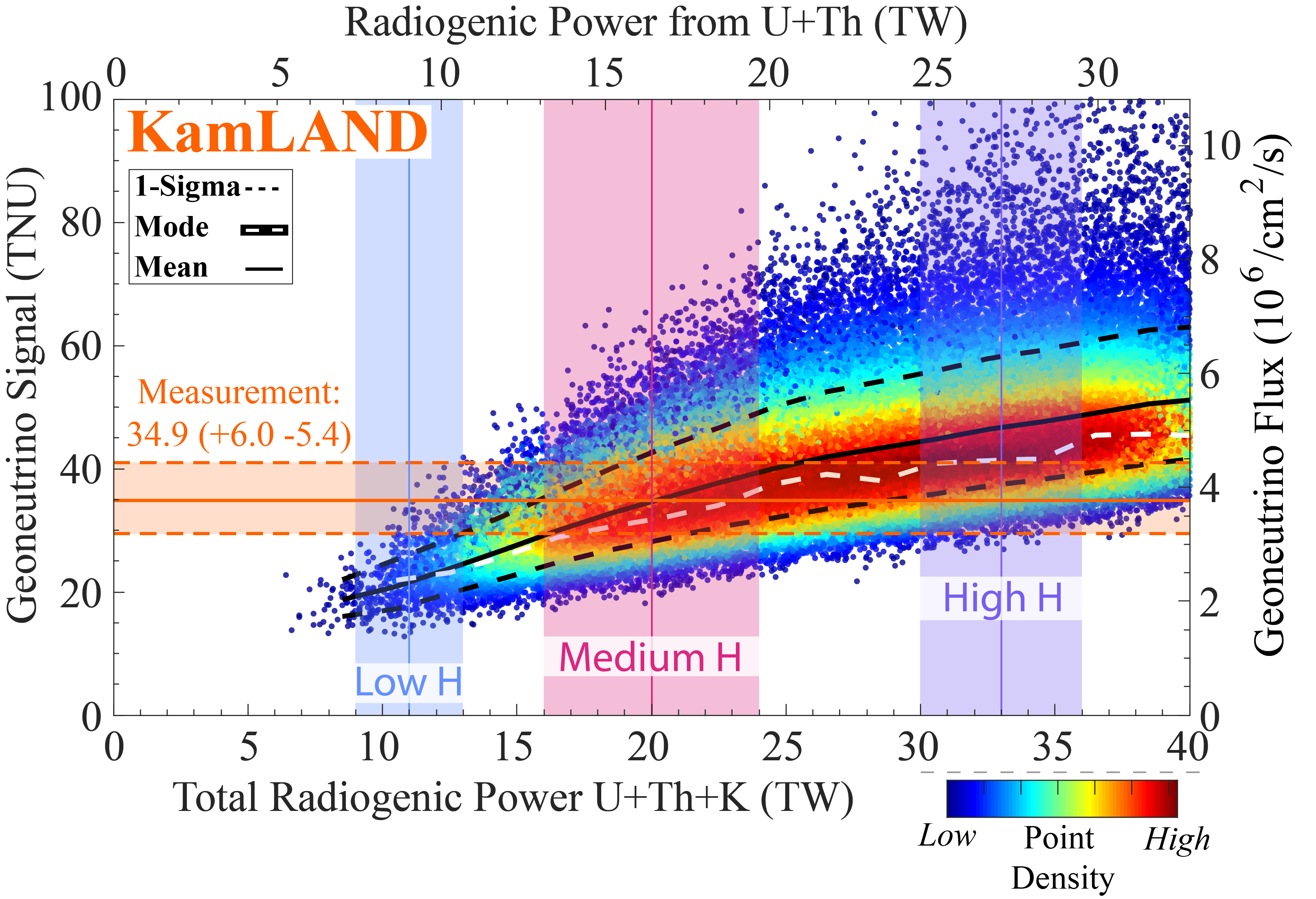}
    \caption{Interpretation of the measured geoneutrino signal at KamLAND from \citeA{watanabe2016} (34.9$^{+6.0}_{-5.4}$ TW). Geoneutrino signal (geoneutrino flux, right y-axis) is plotted against total radiogenic power from U, Th, and K (U + Th only, top x-axis) from the model described in this study. Plotted points represent $\sim$7$\times$10$^4$ Monte Carlo simulation results with BSE radiogenic power varying randomly from 0 to 40 TW. Variability in the y-axis for a given x-value is due to uncertainty in crustal mass and HPE abundances. Models of BSE heat production are overlayed for low-H (11 $\pm$2 TW), medium-H (20 $\pm$4 TW), and high-H (33 $\pm$2 TW). The 1-sigma, mean, and mode are calculated in bins every two TW. The placement of the mode lower in TNU than the mean implies log-normally distributed data. The lack of data points for low BSE TW reflect our assumptions of a minimal amount of HPE within the crust.}
    \label{fig:TNU_TW}
\end{figure}

Antineutrino detectors are presently counting geoneutrinos at KamLAND (Kamioka, Japan; 1 kton) and Borexino (Gran Sasso, Italy; 0.3 kton); future detectors include SNO+ (1 kton; online 2019) in Sudbury, Canada, JUNO (20 kton; online 2022) in Guangdong, China, and Jinping (4 kton; unknown start date) in Sichuan, China. Although the Earth is emitting $\sim10^6$\,$\bar{\nu_e}/$cm$^2/$s, few events are detected annually at KamLAND (∼14/yr) and Borexino (∼4/yr) because of the combined effects of the inverse square law (intensity $\propto$\,1/distance$^2$) and the neutrino's small interaction cross section. Of these detected events, \citeA{araki2005} estimated that 50\% are derived from U and Th in the HPE-enriched upper continental crust within $\sim500$\,km of the detector (known as “near-field crust”). The remaining signal comes equally from the rest of the continental crust (i.e. the “far-field crust”; $>500$\,km) and the mantle. Consequently, the signal from the crust ($\sim40$\,km thick) overpowers that from the more massive mantle ($\sim2900$\,km thick). In order to determine the mantle contribution to the geoneutrino flux --- and therefore the amount of U and Th within the mantle --- it is necessary to have precise and accurate estimates of the crustal contribution. Current measurements from KamLAND and Borexino do not yet allow for discernment between different BSE models (Figure \ref{fig:TNU_TW}).

\begin{figure}
\centering
\includegraphics[width=0.5\linewidth]{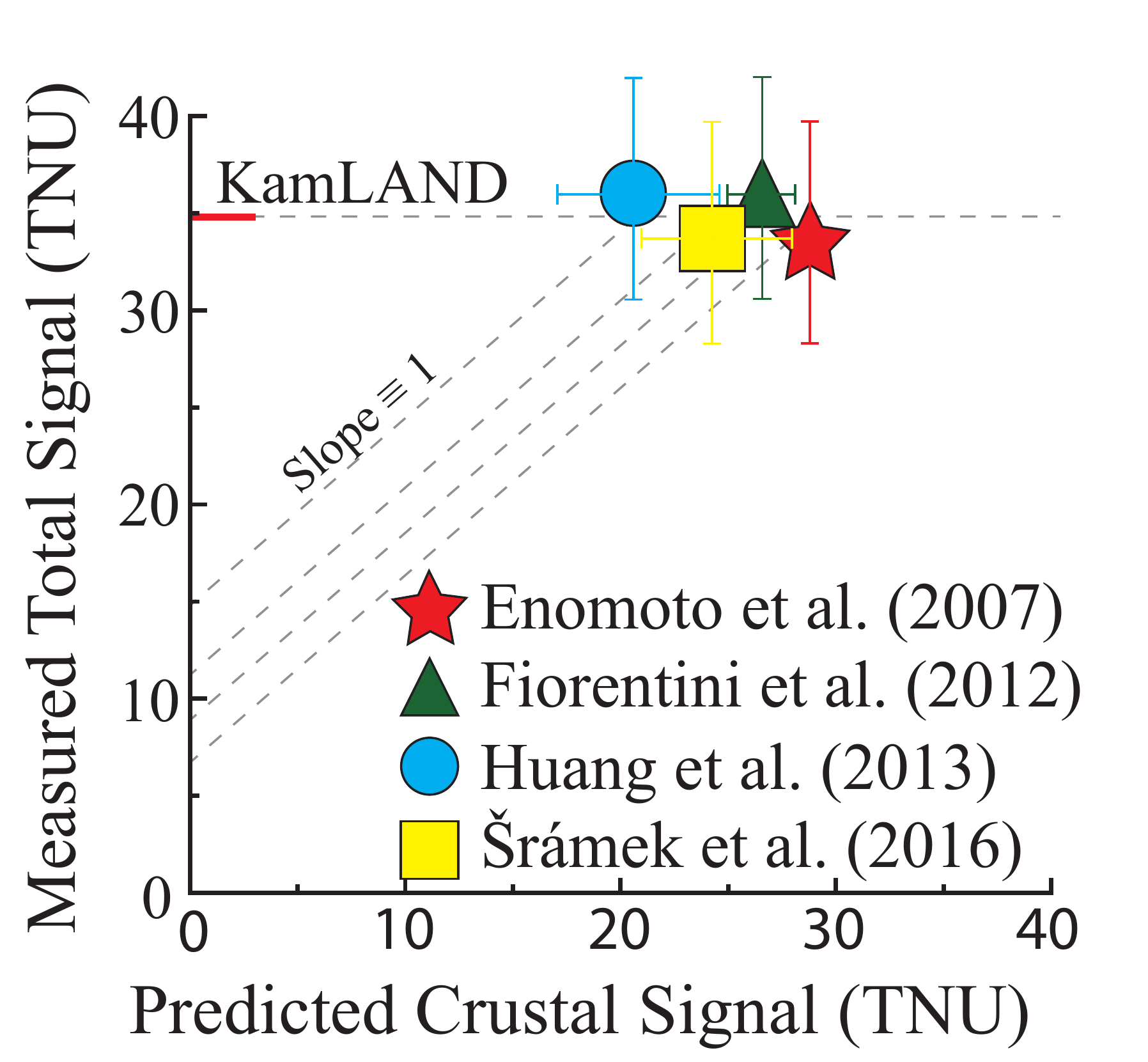}
\caption[Model predictions of geoneutrino signal at KamLAND.]{Discrepancy between model predictions of geoneutrino signal at KamLAND from \citeA{enomoto2007}(28.2 TNU), \citeA{fiorentini2012}(26.5 $\pm$1.52 TNU), \citeA{huang2013}(20.6$^{+4.0}_{-3.5}$ TNU), and \citeA{sramek2016}(24.2 $\pm$3.5 TNU). Values share the same measured total signal (y-axis) from \citeA{watanabe2016} (34.9$^{+6.0}_{-5.4}$ TNU) but are offset for visibility. The y-intercept of a slope 1 line gives the signal from the mantle after removal of the crustal contribution.}
\label{fig:kam_summary}
\end{figure}

Endeavors to understand the crustal signal constitute a major effort in geoneutrino research. Models of the global crust employ a combination of global and regional seismic and gravity data with extrapolation to areas with minimal available data. Similarly, geochemical data from global compilations representative of sediment, upper, middle, and lower crustal layers are combined with these physical models. These joint models make simplifying assumptions regarding the bulk composition of each layer of the continental crust. Only recently did these geochemical model estimates include lateral and vertical spatial variability, although not in all layers of the crust \cite<e.g.,>{huang2013}. A further confounding issue is that there exist different geophysical models of the bulk lithosphere and each yields slightly different geoneutrino estimates. From the variance between these models and lack of direct measurement, it is unclear which model provides a more accurate representation of the crust. For example, estimated signals at KamLAND can vary by as much as 40\%, with particular distinction between models produced respectively by the geoscience and physics communities (Figure \ref{fig:kam_summary}).

This paper aims to update the geoneutrino model of \citeA{huang2013} by application of their geochemical methods to three different geophysical reference models, including a more recent physical model. First, the geophysical structure and properties of the crust from three different geophysical models are described. Geochemical methods are applied to these geophysical models to yield a 3D description of the amount and distribution of [U,Th,K] in the crust. Finally, these combined geophysical and geochemical models are used to calculate the geoneutrino signal at six detector locations. Uncertainties for all input parameters are defined, correlated, and propagated using Monte Carlo methods. Finally, geochemical and geoneutrino signal modeling results are compared and discussed.

\section{Geophysical Model}

\begin{figure}
\centering
\includegraphics[width=0.8\linewidth]{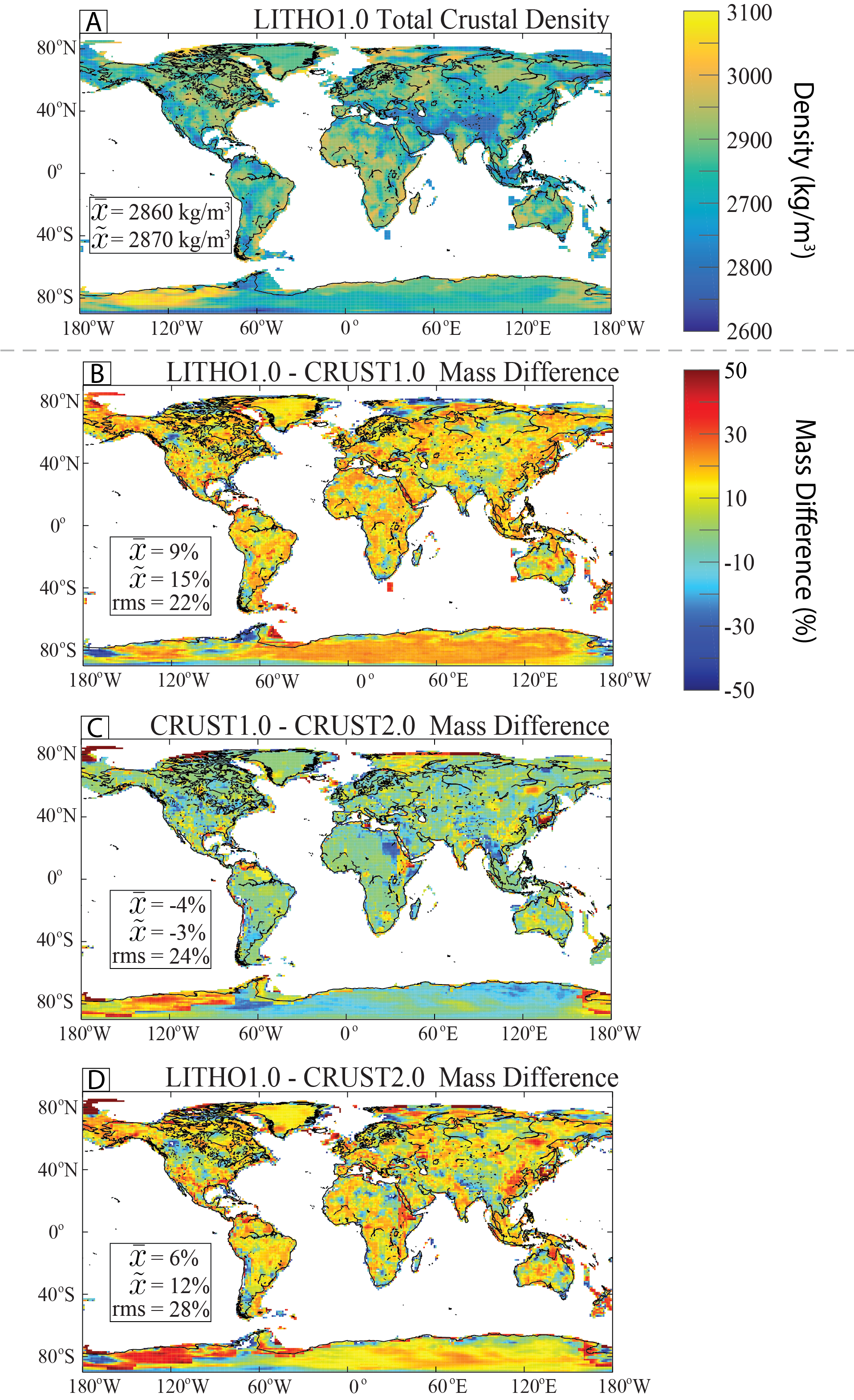}
\caption{Physical descriptions of the continental crust based on geophysical models CRUST2.0, CRUST1.0, and LITHO1.0. Panel A: Density (kg/m$^3$) of LITHO1.0 continental crust within each 1$\times$1 degree cell. The density distribution shown in the figure is consistent with those of CRUST2.0 and CRUST1.0. Panel B, C, and D: Differences ($\%$) in mass of each 1$\times$1 degree cell between LITHO1.0 - CRUST1.0, CRUST1.0 - CRUST2.0, and LITHO1.0 - CRUST2.0. The difference is calculated as (X-Y)/X. Differences in mass seen in panels B, C, and D reflect differences in crustal thickness. In all panels, $\bar{x}$ and $\tilde{x}$ denote the mean and median, respectively. Panels B, C, and D report root mean square (rms), which provides an estimate of the magnitude of values regardless of their sign.}
\label{fig:DensityMassDiff}
\end{figure}

The lithosphere is the outer rigid silicate shell of the Earth and is composed of the thick ultramafic lithospheric mantle and its overlying, relatively thinner and more felsic crust. This crust is either mafic oceanic or intermediate continental crust. In the past few decades authors moved from characterizing the vertical profile of the continental crust as a single bulk layer to a combination of an upper crust and lower crust \cite<e.g.,>{hacker2015}, an upper, middle, and lower crust \cite<e.g., CRUST5.1;>{mooney1998}, or as a continuously changing density structure \cite<e.g., GEMMA1.0;>{reguzzoni2015}. Furthermore, authors have begun to laterally characterize the crust as an assemblage of different groups with similar tectonic and seismic structure, something that has previously been performed \cite<e.g.,>{pakiser1966} but rarely with regard to both lateral and vertical variations. Modern global physical models are constructed using global and local seismic studies, gravity surveys, or a combination of both. The primary utility of global geophysical models is for crustal corrections in mantle tomography \cite{mooney1998} and interpretation of gravity measurements \cite<e.g.,>{reguzzoni2015}. These models also serve as the basis for geochemical and geoneutrino modeling. 

\begin{figure}
\centering
\includegraphics[width=1\linewidth]{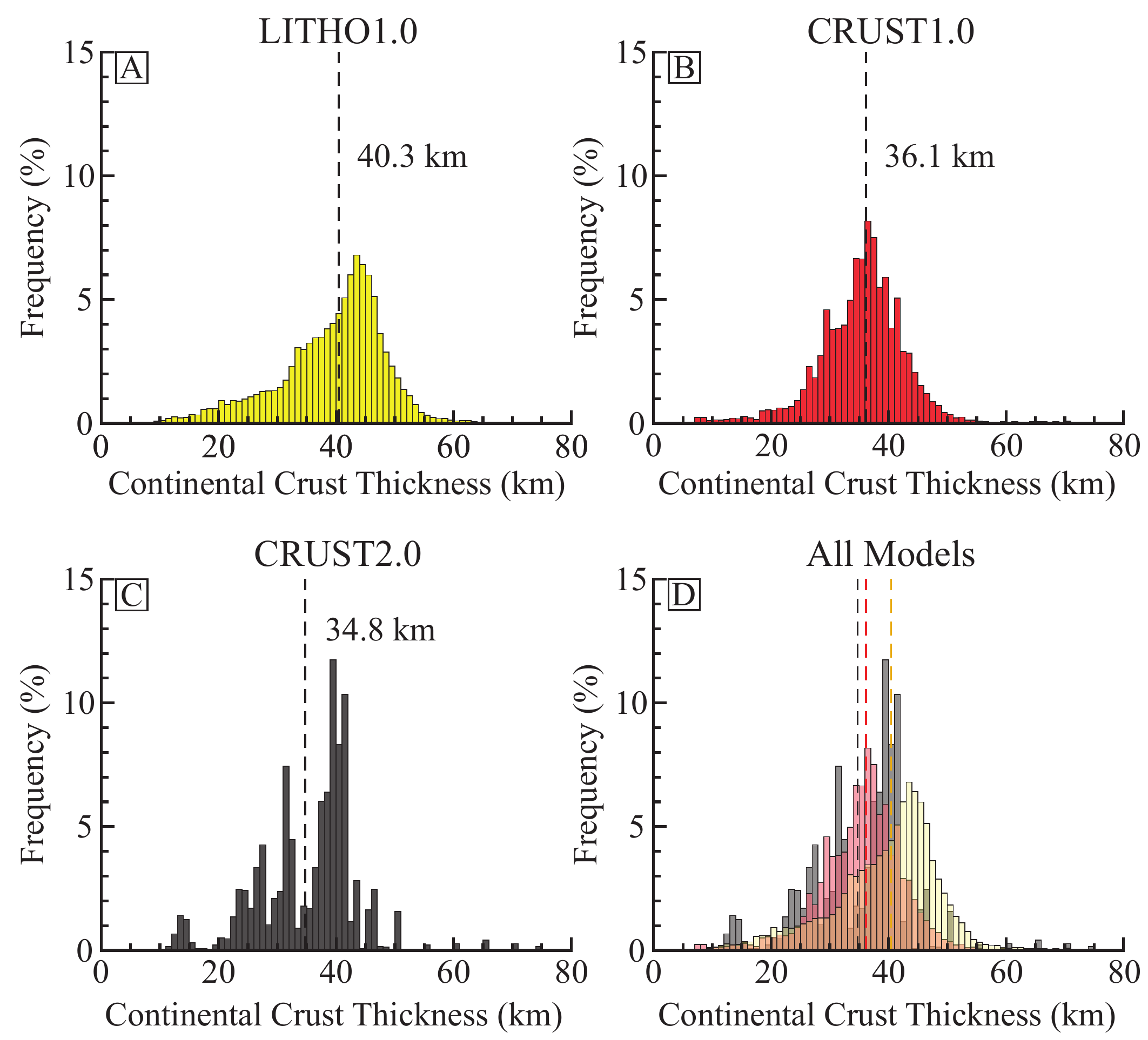}
\caption[Histograms of continental crust thickness.]{Frequency histograms of continental crust thickness for LITHO1.0, CRUST1.0, and CRUST2.0. Panel-D shows the histograms overlapped. Dashed vertical lines and accompanying text mark the surface area weighted average continental crust thickness.}
\label{fig:thick_hists}
\end{figure}

CRUST5.1 \cite{mooney1998} provided seismic compressional ($V_P$) and shear wave ($V_S$) velocity averages for regions with similar crustal structure in order to extrapolate to regions where there was limited or no data \cite{mooney1998}. This model defined the crust as three layers (upper, middle, and lower) with a sediment layer on top. The subsequent family of geophysical models built on CRUST5.1's data and methods have a generational increase in input data, resulting in resolution changes from CRUST5.1 (5\textdegree{}lat $\times$ 5\textdegree{}lon, i.e., 2592 tiles laterally) $\rightarrow$ CRUST2.0 \cite<2\textdegree{} $\times$ 2\textdegree{}, 16200 tiles;>{bassin2000} $\rightarrow$ CRUST1.0 \cite<1\textdegree{} $\times$ 1\textdegree{}, 64800 tiles;>{laske2013}). Of note, CRUST2.0 and CRUST1.0 provide limited transparency of the model inputs or methodology aside from AGU and EGU abstracts, respectively. The most recent iteration of the "CRUST" family of models, LITHO1.0 ($\sim$1\textdegree{} $\times$ 1\textdegree{}) \cite{pasyanos2014}, perturbed CRUST1.0 parameters (density, seismic speeds $V_P$ and $V_S$, layer thickness) to fit a global surface wave dataset and for the first time included lithosphere-asthenosphere boundary (LAB) depths. LITHO1.0 reports $V_P$ and $V_S$ as a continuous probability distribution, while CRUST2.0 and CRUST1.0 report a few discrete velocities within each layer. Recently, \citeA{olugboji2017} showed that LITHO1.0 displays a smaller misfit to high-resolution seismic studies in the United States compared to CRUST1.0. CRUST2.0, CRUST1.0, and LITHO1.0 are adopted as the physical basis for our geoneutrino modeling in order to test their effect on modeled outputs. Model parameters relevant to this study are the layer thickness, density, and $V_P$. These geophysical models are not combined with the more recent GEMMA1.0 (as was done by \citeA{huang2013}) due to ambiguity regarding GEMMA1.0's dependence on the adopted density structure of the crust. However, this also means that we do not have a measure of uncertainty on crustal thickness and must rely on estimates from previous studies (see Section \ref{sec:uncertainty}). A comparison of the mass and thickness of layers within these models is shown in Table \ref{tab:physical}, Figure \ref{fig:DensityMassDiff} (density and mass comparisons), and Figure \ref{fig:thick_hists} (thickness comparisons).

Previous geoneutrino reference models were generally built upon the most recent model available. The first model post the start of geoneutrino measurements at KamLAND was that of \citeA{mantovani2004} built using the Preliminary Reference Earth Model \cite<PREM;>{dziewonski1981}, followed by the models of \citeA{enomoto2007} and \citeA{fiorentini2012} for the geoneutrino signal at KamLAND and Borexino (built on the geophysical model CRUST2.0). \citeA{huang2013} is built on a combination of the physical models CRUST2.0, CUB2.0 \cite{shapiro2002}, and GEMMA1.0 and has been used in the prediction of signals at SNO+ \cite{huang2014,strati2017} and JUNO \cite{strati2015}, among others. Finally, \citeA{sramek2016} calculated the signal at Jinping with a model built upon CRUST1.0 \cite{laske2013}.  

For the purposes of this study LITHO1.0 is re-meshed from its original icosahedron equal-area tesselation (level 7; with $\sim$1\textdegree{} spacing) onto a 1\textdegree{}lat $\times$ 1\textdegree{}lon grid. Similarly, CRUST2.0 is re-meshed into a 1\textdegree{} $\times$ 1\textdegree{}{} grid from its previous 2\textdegree{} $\times$ 2\textdegree{} grid. Re-meshing was necessary to more easily compare modeled outputs. We split the LITHO1.0 and CRUST1.0 tiles into continental and oceanic crust using the crust type characterization from CRUST1.0 and split CRUST2.0 by its own classification. For CRUST1.0 and CRUST2.0, oceanic crust is defined as types A0, A1, B-, V1, Y3 or all A and B types, respectively; these correspond to crustal types labeled "oceanic". The final models contain sediment layers (two for CRUST2.0 and 3 for CRUST1.0 and LITHO1.0), upper, middle, and lower crust, and lithospheric mantle. Sedimentary thickness within the "CRUST" family are from a digitization of energy industry datasets \cite{laske1997}. 

Continental lithospheric mantle LAB depths are provided by LITHO1.0; for CRUST2.0 and CRUST1.0 the LAB is set to 175 $\pm$ 75 km depth following \citeA{huang2013}. Although this study focuses on the lithosphere and continental crust, we include a convecting mantle in order to provide an estimate of the total geoneutrino signal expected at a detector. For the convecting mantle we define the lower 750 km as the Enriched Mantle, which equals $\sim$19\% of the mantle by mass \cite{dziewonski1981,arevalo2013}. PREM densities are adopted for the mantle \cite{dziewonski1981}. The mass of the BSE reported in Table \ref{tab:physical} is a combination of PREM for the convecting mantle and either CRUST2.0, CRUST1.0, or LITHO1.0 for the lithosphere.

\begin{sidewaystable}

\begin{table}[H]
\centering
\def\arraystretch{1.5}
\begin{tabular}{c c c c c c c c c c c c c c}
\multicolumn{3}{c}{\multirow{2}{*}{ }} & \multicolumn{3}{c}{CRUST2.0} &&& \multicolumn{2}{c}{CRUST1.0} &&& \multicolumn{2}{c}{LITHO1.0} \\ 
&Layer & $\sim\rho$ (g/cm$^3$) && d (km)\mbox{*} & M (10$^{21}$ kg) &&& d (km)\mbox{*} & M (10$^{21}$ kg) &&& d (km)\mbox{*} & M (10$^{21}$ kg) \\ \hline
\multicolumn{1}{c}{\multirow{5}{*}{CC}} & Sed & 2.2  && 2.18 $\pm$2.1 & 0.8 $\pm$0.1 &&& 1.64 $\pm$2.2 & 0.7 $\pm$0.1&&& 1.54 $\pm$2.0 & 0.7 $\pm$0.1\\ 
\multicolumn{1}{c}{} 					  & UC     & 2.75 && 11.60 $\pm$3.9 & 7.0 $\pm$0.9 &&&  11.71 $\pm$4.0 & 6.3 $\pm$0.8&&& 12.79 $\pm$4.1 & 6.9 $\pm$0.8\\ 
\multicolumn{1}{c}{} 				      & MC      & 2.84 && 11.18 $\pm$3.4 & 7.2 $\pm$0.9&&&  11.57 $\pm$3.0 & 6.4 $\pm$0.8&&& 13.06 $\pm$3.8 & 7.3 $\pm$0.9\\ 
\multicolumn{1}{c}{}                    & LC      & 3.02 && 9.93 $\pm$2.9 & 6.7 $\pm$0.8&&&  10.73 $\pm$2.7 & 6.2 $\pm$0.8&&& 12.22 $\pm$3.6 & 7.2 $\pm$0.9\\ 
\multicolumn{1}{c}{}                    & Bulk CC & 2.9 && 34.25 $\pm$8.8 & 21.8 $\pm$2.6&&&  35.53 $\pm$7.6 & 19.6 $\pm$2.4&&& 39.60 $\pm$9.1 & 22.2 $\pm$2.6\\ \arrayrulecolor[gray]{0.7}\cline{1-3} \cline{5-7} \cline{9-11} \cline{13-14}
\multicolumn{1}{c}{\multirow{2}{*}{OC}} & Sed 	& 1.9 && 1.86 $\pm$0.2 & 0.3 $\pm$0.0&&&  1.90 $\pm$0.1 & 0.4 $\pm$0.1&&& 1.90 $\pm$0.1 & 0.4 $\pm$0.1\\  
\multicolumn{1}{c}{} 				      & C   	& 2.9 && 6.80 $\pm$1.5 & 5.6 $\pm$0.7&&&  7.82 $\pm$2.9 & 7.1 $\pm$0.9&&&  10.46 $\pm$4.3 & 9.2 $\pm$1.1\\ \arrayrulecolor[gray]{0.7} \cline{1-3} \cline{5-7} \cline{9-11} \cline{13-14}
\multicolumn{1}{c}{\multirow{3}{*}{Mantle}}					  & LM  	& 3.34 && 139.9 $\pm$75 & 102.8 $\pm$53 &&&  139.5 $\pm$75 & 88 $\pm$43 &&& 114.3 $\pm$82 & 63 $\pm$7.5\\ 
\multicolumn{1}{c}{} & DM 					    	& 4.4 && 1966  & 3149.9 &&& 1966  & 3168.5 &&&  1987 $\pm$84 & 3187.3 \\ 
\multicolumn{1}{c}{} & EM 							& 5.4 && 750 & 754 &&& 750 & 754 &&&  750 & 754 \\ \arrayrulecolor[gray]{0.7} \cline{1-3} \cline{5-7} \cline{9-11} \cline{13-14}
\multicolumn{1}{c}{} & BSE 							& 4.45 && 2891 & 4035 &&& 2891 & 4038 &&& 2891 & 4036 \\ \arrayrulecolor{black}\hline
\\ 

\end{tabular}
\caption{Physical properties of the layers of the Bulk-Silicate Earth (BSE) modeled using the geophysical models CRUST2.0, CRUST1.0, and LITHO1.0. The density reported is the mean between the three models with variation between model layers of $<$2.5\% (except 10.5\% variation for lower crust). BSE mass is from a combination of PREM \cite{dziewonski1981} derived mass of the DM and EM combined with the lithosphere mass. \mbox{*} Thicknesses are un-weighted of the 1$\times$1 degree data unlike the area-weighted average reported in Figure~\ref{fig:thick_hists}. }
\label{tab:physical}
\end{table}

\end{sidewaystable}

\section{Geochemical Model}
The Earth's crust is composed of mafic oceanic crust (on average $\sim7$\,km thick) and relatively more felsic continental crust (avg. $\sim35$\,km thick). The geochemistry of the continental crust has been explored extensively over the past century, particularly the accessible upper crust. Generally the crust is thought to decrease in $SiO_2$ with depth from 67\% in the upper crust to 53\% in the lower crust \cite{rudnick2014}. These observations are from surface exposures of the upper, middle, and lower crust as well as xenolith samples. Furthermore, it has long been observed that increasing metamorphic grade cannot account for the increase in the observed $V_P$ with depth, rather requiring a compositional increase in maficity \cite{christensen1995}. The relationship between seismic velocity and chemical composition has been observed empirically, due to the chemical effect on the shear and bulk modulus of which $V_P$ relies \cite{christensen1965,pakiser1966,christensen1995}. However, observed $V_P$ of different lithologies are non-unique, requiring further assumption to have any meaningful result when inverting $V_P$ for composition. The methodology of \citeA{huang2013} is adopted here, which assumes the continental middle crust is composed of amphibolite and the continental lower crust of granulite metamorphic rocks. These assumptions are based on surface exposed crustal cross-sections and xenolith data \cite{rudnick2014}. \citeA{huang2013} correlated $V_P$ and $SiO_2$ and $SiO_2$ and [U, Th, K] abundance of amphibolite and granulite by a linear relationship between mafic and felsic endmembers. This correlation is characterized by a simple mass balance defined by the following equations: 

\begin{equation}
V_{model} = V_f  f + V_m  m
\end{equation}
\begin{equation}
1 = f + m
\end{equation}
\begin{equation}
a_{output} = a_f f + a_m m
\end{equation}

where $V_f$, $V_m$, $a_f$, and $a_m$ are the $V_P$ and abundance of the felsic and mafic end-members of amphibolite (middle crust) and granulite (lower crust), $V_{model}$ is the $V_P$ provided by the geophysical model, and \textit{f} and \textit{m} are the mass proportions of felsic and mafic endmembers. The velocities of the amphibolite and granulite datasets are derived from laboratory measurements at room temperature and 600 MPa \cite{huang2013}. These measurements are temperature corrected ($-4\times10^{-4}$\,km/s/\textdegree{}C) following a mid-range geotherm ($q_s=60$\,mW/m$^2$) \cite{pollack1977,turcotte2014}) and pressure corrected ($2\times10^{-4}$ km/s/MPa) from pressures calculated from the parameters provided by the geophysical model ($\mathrm{d}P = \rho g\mathrm{d}h$) \cite{huang2013}. Temperature and pressure corrections are from empirical studies \cite{christensen1995,rudnick1995}. In cases where the $V_P$ from the geophysical model is larger than the mafic endmember or smaller than the felsic endmember, we set $f$ to be 0 or 1, respectively. In LITHO1.0, this occurs in $\sim$11\% (36\% $f\!=\!0$ and 84\% $f\!=\!1$, proportionally) of middle crustal and $\sim$42\% (98\% $f\!=\!0$ and 2\% $f\!=\!1$, proportionally) of lower crustal continental tiles. Model $V_P$ outside the range of endmembers also occurs in CRUST2.0 and CRUST1.0, with the middle (lower) crust having 11\% (23\%) and 0\% (64\%), respectively. This method is especially problematic for the lower crust, but for the purposes of this study the impact has a negligible effect on the geoneutrino signal as the lower crust is generally depleted in [U,Th,K] and is farther from the geoneutrino detectors than the upper or middle crust (see Section \ref{sec:geo_sig}). 

HPE abundance for the upper crust is not calculated from a $V_P$ and composition correlation because near surface processes strongly affect observed $V_P$ and because we cannot simplify the upper crust as a single metamorphic grade (as we do for the middle and lower crust). Without the simplifying assumption of a single metamorphic grade there are too many non-unique lithologies that could be present. Instead values calculated from a meta-analysis of previously published upper crust estimates is adopted \cite{rudnick2014}, which equates to a sigma-mean value rather than a standard deviation. Although this uncertainty is not consistent with our overall adoption of 1-sigma uncertainties in this study, the analysis of \citeA{rudnick2014} currently provides the best estimate of upper crust composition. 

Sediment compositions are from the GLOSSII model of subducted sediments, as these material would be the remnants of continental weathering and marine activity \cite{plank2014}. This assumption is acceptable as the location of sedimentary basins in the geophysical model is generally along coastlines. The continental lithospheric mantle is characterized by a suite of xenoliths compiled by \citeA{huang2013}. For the oceanic crust we adopt the bulk oceanic crust composition from \citeA{white2014}. The lithospheric mantle below oceanic crust is assumed to have the same composition as the depleted mantle. 

The BSE composition is calculated from the U abundance given by \citeA{mcdonough1995}, whereas Th and K are calculated from observed Th/U and K/U ratios \cite{arevalo2009,wipperfurth2018}. Similarly, a conservative upper mantle (LAB to 750 km above the core-mantle boundary) U abundance is assumed based on the compilation from \citeA{arevalo2013} with the Th and K abundance calculated from observed Th/U and K/U values from mid-ocean ridge basalt samples \cite{arevalo2009,wipperfurth2018}. The abundances within the enriched mantle are the remainder from subtraction of the lithosphere and upper mantle from the BSE composition (Table \ref{tab:geochem}).

\begin{table}[h!]
\centering
\def\arraystretch{1.5} 
\begin{tabular}{c l c c c}
  & & CRUST2.0 & CRUST1.0 & LITHO1.0 \\ \hline
 \multirow{4}{1.5cm}{\centering Upper Crust $^a$} & U ($\mu$g/g)   & & 2.7 $\pm$0.60       & \\
                               & Th ($\mu$g/g)  & & 10.5 $\pm$1.0          &\\
                               &  K (wt\%) & & 2.32 $\pm$0.19           & \\
                               & K/U       & 8,900 $^{+1,600}_{-1,300}$    & 8,900 $^{+1,600}_{-1,300}$ & 8,900 $^{+1,600}_{+1,300}$\\
                               & Th/U      & 4.0 $^{+0.7}_{-0.6}$           & 4.0 $^{+0.6}_{-0.6}$        & 4.0 $^{+0.7}_{-0.6}$\\ 
                               & P (TW)    & 4.2 $^{+0.9}_{-0.8}$           & 3.8 $^{+0.8}_{-0.7}$      & 4.1 $^{+0.9}_{-0.7}$\\ \hline       

 \multirow{4}{1.5cm}{\centering Middle Crust $^b$} & U    & 0.82 $ ^{+0.87}_{-0.42} $ & 0.94 $ ^{+0.95}_{-0.47} $ & 0.84 $ ^{+0.87}_{-0.43} $ \\
                               & Th   & 3.64 $ ^{+6.27}_{-2.30} $ & 4.62 $ ^{+7.11}_{-2.80} $ & 3.88 $ ^{+6.03}_{-2.36} $ \\
                               & K   & 1.42 $ ^{+1.52}_{-0.73} $ & 1.69 $ ^{+1.66}_{-0.84} $ &  1.47 $ ^{+1.47}_{-0.74} $ \\ 
                               & K/U       & 14,700 $\pm$2,200 & 15,300 $\pm$2,000 & 14,900 $\pm$2,000\\
                               & Th/U      & 4.6 $\pm$1.4           & 5.0 $\pm$1.2 & 4.7 $\pm$1.2\\ 
                               & P    & 1.4 $^{+1.8}_{-0.8}$           & 1.5 $^{+1.7}_{-0.8}$        & 1.5 $^{+1.7}_{-0.8}$\\ \hline 

 \multirow{4}{1.5cm}{\centering Lower Crust $^c$}  & U    & 0.15 $ ^{+0.22}_{-0.09} $ & 0.19 $ ^{+0.27}_{-0.11} $ & 0.17 $ ^{+0.24}_{-0.10} $ \\
                               & Th   & 0.81 $ ^{+1.94}_{-0.57} $ & 1.14 $ ^{+2.89}_{-0.82} $ & 0.95 $ ^{+2.27}_{-0.67} $\\
                               & K  & 0.70 $ ^{+0.80}_{-0.37} $ & 0.91 $ ^{+1.12}_{-0.50} $ & 0.78 $ ^{+0.90}_{-0.42} $ \\
                               & K/U       & 39,700 $\pm$10,600    & 42,000 $\pm$10,400 & 40,700 $\pm$10,400\\
                               & Th/U      & 5.3$^{+2.5}_{-1.7}$           & 6.0 $^{+3.1}_{-2.1}$        & 5.6 $^{+2.6}_{-1.8}$\\ 
                               & P    & 0.4$^{+0.6}_{-0.2}$           & 0.5$^{+0.8}_{-0.3}$        & 0.5$^{+0.7}_{-0.3}$\\ \hline \hline          

 \multirow{6}{2cm}{\centering Bulk Continental Crust}   & U  & 1.31$^{+0.35}_{-0.28}$   & 1.35$^{+0.36}_{-0.29}$     & 1.29$^{+0.35}_{-0.27}$ \\
                               & Th   & 5.81$^{+1.96}_{-1.47}$     & 6.25$^{+2.26}_{-1.66}$     & 5.77$^{+1.99}_{-1.48}$\\
                               & K  & 1.81$^{+0.51}_{-0.39}$        & 1.96$^{+0.57}_{-0.44}$     & 1.81$^{+0.52}_{-0.40}$\\ 
                               & K/U       & 11,500$^{+1,800}_{-1,600}$    & 12,100$^{+2,000}_{-1,700}$ & 11,800$^{+2,000}_{+1,700}$\\
                               & Th/U      & 4.4$^{+0.7}_{-0.6}$           & 4.6$^{+0.8}_{-0.7}$        & 4.5$^{+0.8}_{-0.7}$\\ 
                               & P     & 7.0$^{+2.1}_{-1.6}$           & 6.6$^{+2.0}_{-1.6}$        & 7.0$^{+2.1}_{-1.6}$\\ \arrayrulecolor{black} 
                               \\
\end{tabular}
\caption{Calculated abundances, element ratios, and power for the upper crust, middle crust, lower crust, and  bulk-continental crust. Upper crust abundances are same in all models, but power varies as mass of crust varies. Reported uncertainties are 1$\sigma$. Bulk continental crust values are weighted by mass of each layer and include the sediment contribution (see Supplementary Information for sediment values). K/U and Th/U are mass ratios.\\
a = Values from \citeA{rudnick2014}. \\
b = Derived from $V_P$--(U,Th,K) relationship of amphibolites.\\
c = Derived from $V_P$--(U,Th,K) relationship of granulites.}
\label{tab:geochem}
\end{table}

\section{Uncertainties and Correlation} \label{sec:uncertainty}

The attribution of uncertainty is the most difficult and time consuming part of numerical modeling. When available, uncertainties directly reported by the data source are adopted. When not available, relative uncertainties in agreement with literature estimates are adopted and correspond to 1$\sigma$, except for abundances in the upper crust. [U,Th,K] abundances display log-normal distributions, a trait shared with other incompatible elements \cite{ahrens1954,mcdonough1990,wipperfurth2018}. Model outputs, which depend on asymmetrical abundance distributions, are reported as geometric mean $\pm$ 1$\sigma$ uncertainties. 5\% uncertainty is ascribed to $V_P$, which is primarily based on the recent comparison of a high-resolution surface wave model of the US with CRUST1.0 and LITHO1.0 \cite{olugboji2017}, although this is in agreement with earlier estimates \cite{mooney1998}. Because density is derived from $V_P$ in the "CRUST" family of models, 5\% uncertainty is also applied to density. 12\% uncertainty is ascribed to crustal thickness based on the comparison of CRUST2.0, CUB2.0, and GEMMA1.0 by \citeA{huang2013}. Uncertainty on mantle thickness are not incorporated in order to decrease computation time, although the effect on geoneutrino signal is expected to be negligible as the uncertainty on the mantle mass is $\sim0.5\%$. 

The correlation of uncertainties is equally as important as absolute uncertainty attribution and largely dictates the relative magnitude of the uncertainty on the modeling output. For example, no correlation in thickness of layers, and thereby no correlation of the mass, results in negligible uncertainty on the estimated mass of the crust ($\ll$1\%). This result is unrealistic as we do not know the crustal mass to such precision. Thickness is correlated vertically across all layers and laterally across all 1\textdegree{} $\times$ 1\textdegree{} tiles; in this way the crust is modeled when it is most massive (all layers as thick as possible) and least massive (all layers as thin as possible). A partial uncertainty would be ideal but is impossible without in-depth knowledge of the correlation of input data used to construct the geophysical model (e.g., the foundational data of LITHO1.0 and their correlations). Neither $V_P$ nor density are correlated between the middle and lower crust as it is unclear how $V_P$ and density would shift given a change in one of the layers. Similar to thickness, $V_P$ and density are laterally correlated within a layer. Abundances are correlated within each voxel as there is a longstanding observation of a general correlation between [U,Th,K] due to their similar incompatibilities during mantle and crustal melting. This correlation also applies to the endmember abundances for amphibolite and granulite. However, abundances are not correlated vertically between layers because the samples used to estimate the abundances of the upper, middle, and lower crust are not the same samples and therefore variability within each dataset is independent of the others. There is, however, an inherent correlation between the middle and lower crust in our model because of the correlation of $V_P$ within these layers from the geophysical model coupled with our geochemical methods. Abundances are correlated laterally within a layer, which coupled with lateral correlations of $V_P$, density, and thickness results in models when layers have the most massive and least massive amount of HPE possible. All correlations described above assume 100\% correlation between parameters. Our correlation approach is conservative; removal of lateral correlations would significantly decrease geoneutrino signal uncertainties but is not obviously justified at this time.

\section{Numerical Model and Geoneutrino Signal} \label{sec:geo_sig}
The geophysical and geochemical information are combined into a coherent model, with attributes assigned for each 1\textdegree{} $\times$ 1\textdegree{} cell for each layer. A Monte Carlo simulation is used to propagate uncertainty with 3$\times$10$^4$ iterations, as at this level we reach stability of outputs. In each iteration the input variable's probability density function (PDF) is randomly sampled. If the parameter is correlated with another, then both parameters are sampled from the same part of the PDF. This method allows for easy combination of normal and log-normal distributions. The code is written for parallel computing in MATLAB using either LITHO1.0, CRUST1.0, or CRUST2.0, taking $\sim$8 hours on a 4 core (8 thread) Intel CPU for $3\times10^4$ iterations of the Monte Carlo. In this way we have provided an environment to use and test the geoneutrino response given different geophysical models within a self-consistent framework. The effect of the number of iterations on the final signal was tested by performing 10 repeated calculations with $20\times10^4$ iterations each, with relative variation in the central value and 1-sigma uncertainties of 0.1\% and 0.5\%, respectively.

For each voxel (3D pixel) within a layer, the mass was calculated from the the density and thicknesses provide by the geophysical model. Multiplication of the mass by HPE abundance yields the mass of HPE within each voxel. The heat production is calculated using updated decay and heat output parameters from \citeA{ruedas2017}. Finally, the geoneutrino signal can be calculated for U and Th. Iterations where the enriched mantle is more depleted in HPE abundances than the depleted mantle, which occurs in $\sim$25\% of iterations, are discarded. This avoids conflicts with observed OIB samples with enriched HPE abundances compared to MORB samples \cite{arevalo2013} as well as avoiding an enriched mantle with no HPE.

\subsection{Calculating Signal}
We calculate the predicted geoneutrino signal at a specified detector location defined by latitude, longitude, and depth below surface (see Supplementary Information). The energy spectrum of detected antineutrinos is written \cite<e.g.,>{dye2012} as

\begin{equation} \label{eq:flux}
\frac{dN(E_{\bar{\nu}_e},\vec{r})}{dE_{\bar{\nu}_e}} = \epsilon\frac{N_A\lambda}{\mu}\sigma_P(E_{\bar{\nu}_e}) \frac{dn(E_{\bar{\nu}_e})}{dE_{\bar{\nu}_e}}\int_{\Earth}^{}P_{ee}(E_{\bar{\nu}_e},|\vec{r}-\vec{r}\prime|)d\vec{r}\prime\frac{a(\vec{r}\prime)\rho(\vec{r}\prime)}{4\pi|\vec{r}-\vec{r}\prime|^2}
\end{equation}
where the meaning of symbols and their units are explained in Table \ref{tab:physics_params}. \\

\begin{table}[H]
 \centering
 \def\arraystretch{1.4}
\begin{tabular}{c  c  c}
\textbf{Symbol} & \textbf{Description} & \textbf{Units} \\ \hline
	$\frac{dN(E_{\bar{\nu}_e},\vec{r})}{dE_{\bar{\nu}_e}}$ & ${\bar{\nu}_e}$ detection spectrum & ${\bar{\nu}_e}$  \\
	$\epsilon$ & $10^{32}$ proton $\times$  3.154$\times10^7$ s $\times$ 100\% & proton$\times$s \\ 
	$N_A$ & Avagodro's number & $\frac{atom}{mol}$ \\ 
	$\lambda$ & Decay constant & $\frac{decay}{s\times atom}$ \\ \
	$\mu$ & Atomic mass in kg & $\frac{kg}{mol}$ \\ \
	$\sigma_P(E_{\bar{\nu}_e})$ & ${\bar{\nu}_e}$ cross-section in $m^2$ (function of $E_{\bar{\nu}_e}$) & $\frac{m^2}{proton}$ \\ 
	$\frac{dn(E_{\bar{\nu}_e})}{dE_{\bar{\nu}_e}}$ & ${\bar{\nu}_e}$ emission spectrum & $\frac{{\bar{\nu}_e}}{decay}$ \\ 
	$P_{ee}(E_{\bar{\nu}_e},|\vec{r}-\vec{r}\prime|)$ & Oscillation probability (function of $E_{\bar{\nu}_e}$) & unit-less \\ 
	$a(\vec{r}\prime)$ & Abundance of radionuclide in cell & $\frac{kg}{kg}$ \\ 
	$\rho(\vec{r}\prime)$ & Density of rock in cell &  $\frac{kg}{m^3}$ \\
	$|\vec{r}-\vec{r}\prime|$ & Distance from cell to detector & m \\ \hline
	\\
\end{tabular}
\caption{Variables used in the calculation and propagation of the geoneutrino signal.}
\label{tab:physics_params}
\end{table}

This equation accounts for the number of decays of each isotope, the emitted neutrino energy spectrum, the distance the isotope is from the detector, and the probability of a neutrino oscillation over that distance. The oscillation parameter $P_{ee}$ is calculated from the most recent values from \cite{capozzi2017}. The signal is calculated and propagated for every 75 keV across the IBD-detectable geoneutrino emission spectrum ranging from $\sim1800$~keV to $\sim3300$~keV, as calculated by \citeA{enomoto:nuspec}. The output from equation \ref{eq:flux} is in TNU.

As the geoneutrino signal is dependent on the distance from the detector in inverse square ($|\vec{r}-\vec{r}\prime|^2$) it is necessary during modeling to increase the mesh resolution for tiles near the detector, otherwise the signal would be calculated from only the center of each 1\textdegree{} $\times$ 1\textdegree{} voxel. The mesh resolution is increased until the final signal reaches stability. This results in modeled cell sizes increasing from $\sim$ 100 m$^3$ next to the detector, 10 km$^3$ at 500 km distance, and the initial cell size after 1000 km. The re-meshing is a necessary step in the calculation of the geoneutrino signal but is computationally intensive, increasing computation time by $\sim$40\%.

\section{Results and Discussion}

This study provides an updated reference model to \citeA{huang2013} for the abundance of heat producing elements in the lithosphere, their heat production, the geoneutrino signal at current and future detectors, and a comparison of these outputs from three commonly used geophysical models (Table \ref{tab:FluxCrust}). The crustal abundances and geoneutrino signal are a result of the complex amalgamation of laterally and vertically variable seismic velocity, layer thickness, layer density, and [U,Th,K] abundance. These parameters change from cell to cell in a convoluted way, making analysis of the specific cause of differences in the geoneutrino signal and crustal abundances between geophysical models difficult. For these reasons the following section will attempt to convey some of the major differences when using different geophysical models. A broad overview of different lithospheric thickness models is provided by \citeA{steinberger2016}, which complements this study. 

\subsection{Geoneutrino Signal}
 The geoneutrino signal results at six detectors are reported in Table \ref{tab:FluxCrust}, which includes signal from the bulk continental crust and total expected signal from the BSE. A full breakdown of each signal is included in the Supporting Information. This study finds that the near-field contributes $\sim$40\% of the signal at the detector, with the remaining signal from the rest of the crust ($\sim$35\%) and the mantle ($\sim$25\%). To illustrate the similarity between signals when using CRUST2.0, CRUST1.0, and LITHO1.0, we include in Table~\ref{tab:FluxCrust} the maximum difference in the central value as well as the Overlapping Coefficient (OVL), which measures the degree with which the distributions overlap \cite{inman1989}. 
In our case the OVL is calculated from

\begin{equation} 
\label{eq:ovl}
    OVL = \int min[f_{M_1}(S), f_{M_2}(S)] \mathrm{d}S
\end{equation}

where $f_{M_1}$ and $f_{M_1}$ are normalized probability density functions for geoneutrino signal $S$ calculated with Earth models $M_1$ and $M_2$, and $M_i$ is selected from CRUST2.0, CRUST1.0, or LITHO1.0. The OVL can only reach values from 0 (no overlap) to 1 (complete overlap). The results of this study show that the choice of global physical model has only a small effect on the predicted crustal geoneutrino signal (Table \ref{tab:FluxCrust}). Estimates for the crustal geoneutrino signal at KamLAND, SNO+, and Jinping show larger variation between predictions than the other models (15\% maximum variation in the central value). OVL values are always greater than 70\% overlap, with average overlap greater than 90\%. The OVL values suggests more similarity between LITHO1.0 and CRUST1.0 compared to CRUST2.0. This is in contrast to the apparent similarity in regional masses between CRUST2.0 and CRUST1.0 observed in Figure \ref{fig:DensityMassDiff}.

\begin{sidewaystable}
    
\begin{table}[H]
    \centering
     \def\arraystretch{1.5}
     \setlength{\tabcolsep}{4pt} 
    \begin{tabular}{l c c c c c c c cl}

  &   & \multicolumn{3}{c}{\textbf{Geoneutrino Signal (TNU)}} &   & \multicolumn{3}{c}{\textbf{Overlapping Coefficient (\%)}} \\
\multicolumn{1}{l}{\textbf{Detector}} &   & \multicolumn{1}{c}{\textbf{CRUST2}} & \multicolumn{1}{c}{\textbf{CRUST1}} & \multicolumn{1}{c}{\textbf{LITHO1}} & \boldmath{}\textbf{$\mid\Delta\mid$}\unboldmath{}(\%) & \textbf{L1,C1} & \textbf{L1,C2} & \textbf{C1,C2} \\ \hline

\multirow{2}[0]{*}{KamLAND} & \multicolumn{1}{l}{\textcolor[gray]{0.7}{Bulk CC}} & 22.7$^{+5.9}_{-4.7}$ & 24.2$^{+6.7}_{-5.2}$ & 26.4$^{+7.1}_{-5.6}$ & \textcolor{red}{15} & 85 & \textcolor{red}{74} & 89 \\
  & \multicolumn{1}{l}{\textcolor[gray]{0.7}{Total}} & 34.7$^{+5.8}_{-5.0}$ & 36.6$^{+6.5}_{-5.5}$ & 37.9$^{+6.6}_{-5.6}$ & 9 & 92 & \textcolor{red}{78} & 87 \\\arrayrulecolor[gray]{0.7}\hline
  
\multirow{2}[0]{*}{Borexino} &   & 30.5$^{+8.1}_{-6.4}$ & 29.9$^{+8.0}_{-6.3}$ & 30.5$^{+7.7}_{-6.2}$ & 2 & 96 & 98 & 96 \\
  &   & 43.2$^{+8.0}_{-6.7}$ & 42.9$^{+7.9}_{-6.6}$ & 42.1$^{+7.3}_{-6.2}$ & 1 & 95 & 94 & 98 \\\arrayrulecolor[gray]{0.7}\hline
  
\multirow{2}[0]{*}{SNO+} &   & 37.3$^{+10.2}_{-8.0}$ & 32.9$^{+9.6}_{-7.4}$ & 33.8$^{+9.6}_{-7.5}$ & \textcolor{red}{13} & 95 & 84 & 80 \\
  &   & 49.8$^{+9.7}_{-8.1}$ & 45.7$^{+9.3}_{-7.7}$ & 46.8$^{+9.4}_{-7.8}$ & 8 & 95 & 86 & 82 \\\arrayrulecolor[gray]{0.7}\hline
  
\multirow{2}[0]{*}{JUNO} &   & 28.1$^{+7.5}_{-5.9}$ & 28.0$^{+7.7}_{-6.1}$ & 29.2$^{+8.0}_{-6.3}$ & 4 & 93 & 93 & 99 \\
  &   & 40.5$^{+7.4}_{-6.3}$ & 40.7$^{+7.6}_{-6.4}$ & 40.4$^{+7.4}_{-6.3}$ & 1 & 98 & 99 & 99 \\\arrayrulecolor[gray]{0.7}\hline
  
\multirow{2}[0]{*}{Jinping} &   & 42.5$^{+11.5}_{-9.1}$ & 47.2$^{+12.7}_{-10.0}$ & 48.5$^{+13.1}_{-10.3}$ & \textcolor{red}{13} & 96 & \textcolor{red}{78} & 83 \\
  &   & 55.0$^{+10.9}_{-9.1}$ & 59.9$^{+12.1}_{-10.1}$ & 59.9$^{+12.2}_{-10.1}$ & 9 & 100 & 81 & 81 \\\arrayrulecolor[gray]{0.7}\hline
  
\multirow{2}[0]{*}{Hawaii} &   & 2.3$^{+0.7}_{-0.5}$ & 2.1$^{+0.6}_{-0.5}$ & 2.3$^{+0.7}_{-0.5}$ & 6 & 90 & 93 & 83 \\
  &   & 12.9$^{+2.8}_{-2.3}$ & 12.9$^{+2.8}_{-2.3}$ & 12.8$^{+2.7}_{-2.3}$ & 1 & 98 & 99 & 99 \\\arrayrulecolor[gray]{0.7}\hline

    \\
    \end{tabular}
    \caption{Continental crust (top row) and total (bottom row) signal (in TNU) using each detector location from each geophysical model. $\mid\Delta\mid$ is the maximum difference between the largest and smallest signal's central value estimated at each detector, with $\mid\Delta\mid$ = $\mid$(x-y)/((x+y)/2)$\mid \times$100\%. The overlapping coefficient refers to the degree of overlap of the two distributions listed at the head of each column (from 0 to 100\% overlap; eqn.\ref{eq:ovl}), with L1 = LITHO1.0, C1 = CRUST1.0, and C2 = CRUST2.0. Values of $\mid\Delta\mid$ ($>$10) and overlapping coefficient ($<$80) which show large differences to the other models are highlighted in red.}
    \label{tab:FluxCrust}
\end{table}

\end{sidewaystable}

\begin{figure}
\centering
\includegraphics[width=1\linewidth]{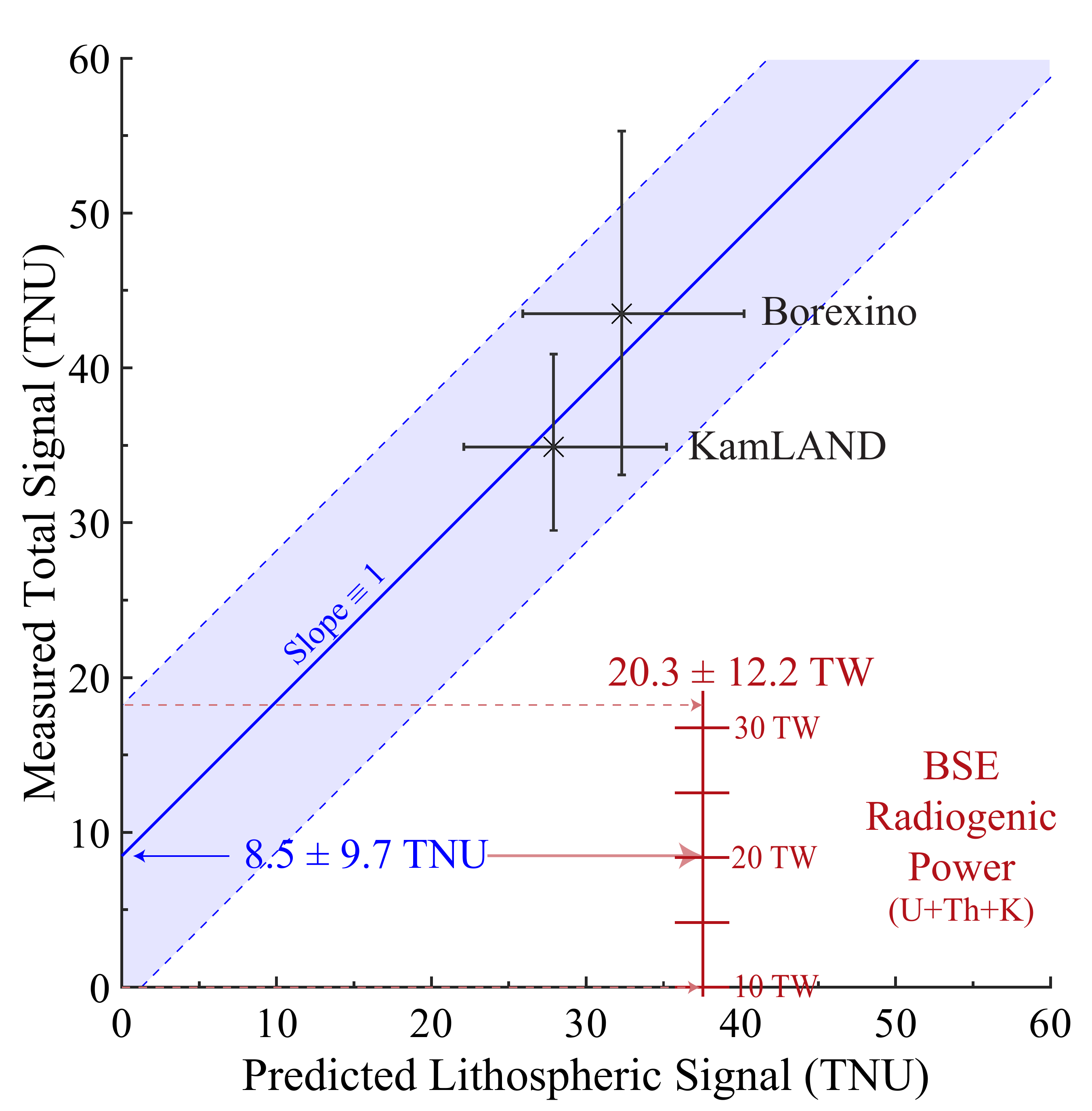}
\caption{Conversion of measured and predicted geoneutrino signals at KamLAND and Borexino to BSE radiogenic power. The slope of the best fit line (blue diagonal) is set to 1. The y-intercept is equal to the mantle component of the geoneutrino signal. X-axis values are from this study using LITHO1.0. Uncertainty on the y-intercept and BSE radiogenic power are reported as 1-sigma.}
\label{fig:slope1}
\end{figure}

The measured signal at Borexino \cite<$43.5^{+11.8}_{-10.4}$ TNU;>{agostini2015} and KamLAND \cite<$34.9^{+6.0}_{-5.4}$ TNU;>[]{watanabe2016} are plotted against the predicted lithospheric signal from LITHO1.0 in Figure \ref{fig:slope1} in order to interpret the observed signals for BSE radiogenic power. Figure \ref{fig:slope1} represents a two-component system whereby the total measured signal (from the BSE) is comprised of the lithospheric and mantle signals. The slope is therefore defined equal to unity (i.e., y = x + b). The y-intercept, which represents the signal from the mantle, is calculated using Monte Carlo methods in order to account for asymmetric X and Y uncertainties. The predicted lithospheric signal for KamLAND and Borexino are correlated during the y-intercept Monte Carlo as most parameters are correlated within the geoneutrino reference model. Conversion of mantle TNU signal to BSE radiogenic power uses the relationship derived using the geoneutrino prediction model (similar to Figure~\ref{fig:TNU_TW} which shows total geoneutrino signal vs. radiogenic power in BSE). The estimated BSE radiogenic heat production of $20.3 \pm12.2$ TW is consistent within 1-sigma of the low, medium, and high-H BSE models with the central value coincident with the medium-H model ($20 \pm4$ TW). Furthermore, the prediction given here using LITHO1.0 is consistent with that predicted by the \citeA{sramek2016} model ($21\pm10$ TW) when using the updated \citeA{watanabe2016} KamLAND measurement (personal communication). Based on Figure \ref{fig:slope1}, an observed mantle signal of 8.5 TNU with uncertainty below some $\pm6$ TNU is needed to discriminate between the different BSE compositional models. This uncertainty corresponds to a 40\% reduction in the measured total and predicted lithospheric signal uncertainty at KamLAND and Borexino.

\subsection{HPE Abundance and Heat Production}

Abundances described in Table \ref{tab:geochem} for all layers are comparable (within 1$\sigma$) to both \citeA{rudnick2014} and \citeA{huang2013}. However, uncertainties reported in the middle and lower crust are larger than those reported by \citeA{huang2013}, likely due to differences in how these values were correlated. K/U and Th/U of the layers of the crust and bulk crust are also in agreement at 1$\sigma$ with previous estimates \cite{huang2013,rudnick2014,wipperfurth2018}.

The bulk-continental crust heat production is negligibly different between the geophysical models (Table \ref{tab:geochem}) and is consistent with previous estimates \cite{mareschal2013,huang2014, rudnick2014}. Furthermore, an estimate of bulk continental heat production of 7.0$^{+2.1}_{-1.6}$ TW (using LITHO1.0) encompasses average heat production in stable Precambrian (5.9 $\pm$0.6 TW) and Phanerozoic (8.3 $\pm$1.0 TW) crust estimated from heat flow studies \cite{mareschal2013}. The similarity of continental crust heat production when using CRUST2.0 and LITHO1.0 is in contrast with the crustal signal similarity of CRUST1.0 and LITHO1.0 (Table \ref{tab:FluxCrust}) or the apparent crustal mass similarity of CRUST2.0 and CRUST1.0 observed in Figure \ref{fig:DensityMassDiff} panel C. The similarity in crustal heat production of CRUST2.0 and LITHO1.0 is simply a reflection of their similar bulk crustal mass compared to the lower mass of CRUST1.0 (Table \ref{tab:physical}).

\subsection{Uncertainties}
The reported uncertainties for crustal abundances of the HPE in Table \ref{tab:geochem} are larger than that reported in \citeA{huang2013}, primarily as a result of increased correlation of abundances and geophysical uncertainties. Increased correlation causes an inverse effect on uncertainty for the bulk crustal K/U and Th/U, resulting in less uncertainty than described by \citeA{huang2013}. Some previous geoneutrino signal predictions did not include uncertainty on geophysical inputs, including crustal thickness \cite<e.g.,>{enomoto2007,fiorentini2012,sramek2016}, meaning these studies report smaller uncertainty on the geoneutrino signal. Similar to signal, the uncertainty on continental crustal heat production is $\sim$50\% larger than that estimated by \citeA{huang2013} ($6.8^{+1.4}_{-1.1}$~TW). In general, geophysical uncertainties (on thickness, density, and $V_P$) account for $\sim30\%$ of the signal uncertainty, with the remaining proportion ($\sim70\%$) from the geochemical inputs. When only considering the geophysical uncertainty, as the same geochemical method was applied to each geophysical model, the calculated crustal signals are still negligibly different.

\section{Future Developments}

The results from this study highlight the convergence of our understanding of the physical nature of the continental crust and the negligible effect of geophysical model on the predicted geoneutrino signal. Due to the strong effect of geochemical variability and distance from the detector on the predicted geoneutrino signal, future geoneutrino modeling needs to focus on better understanding of geochemical variability and a focus on near-field modeling. Below we highlight a few areas which require further exploration in order to better understand the crustal component of the geoneutrino signal.

\subsection{Upper Crust Geochemical Uncertainty}
The attribution of uncertainty on the concentration of heat producing elements in the upper crust (sigma-mean) in most geoneutrino studies (including this one) is not consistent with uncertainty on other parameters (1-sigma), including abundances in other layers. Often cited studies on the upper crust composition report sigma-mean \cite{rudnick2014} or Median Absolute Deviation \cite<MAD;>[]{gaschnig2016}, both of which report smaller error estimates than the standard deviation. Any regional or global geoneutrino modeling should adopt consistent error estimators, be it standard deviation or MAD, or justify the treatment of the upper crust separately from other portions of the crust. Upper crust U abundance used in this study is 2.7 $\pm$ 0.6 $\mu g/g$ (sigma-mean) from \citeA{rudnick2014}, similar to that from \citeA{gaschnig2016} of 2.66 $\pm$ 0.87 $\mu g/g$. The standard deviation of the glacial diamictite data (assumed to sample large portions of the upper crust) from \citeA{gaschnig2016} is 150\% of the MAD. Because the upper crust is the dominant heat and geoneutrino emitter, consistent uncertainty estimates on abundances could significantly increase uncertainties on the predicted crustal geoneutrino signal. For example, attribution of abundances in the upper crust using 1-sigma values from a log-normal fit of \citeA{gaschnig2016} data yields a total signal at KamLAND of 36.4 $^{+9.8}_{-7.7}$ TNU (compared to 37.9 $^{+6.6}_{-5.6}$ TNU reported in Table \ref{tab:FluxCrust}) using LITHO1.0. Use of 1-sigma uncertainty for the upper crust abundances therefore results in an increase of uncertainty of $\sim$40\% in the predicted signal, an observation consistent with other studies \cite<e.g.,>{takeuchi2019}.

\subsection{$V_P$ and Density under Himilayas and Andes}
Future studies with the goal to update LITHO1.0 should aim to better understand the density and seismic structure beneath the Himalayas and Andes regions. The 'CRUST' family of models predicts felsic-like densities in the entire crustal column in these regions (including the expected mafic lower crust; Figure \ref{fig:DensityMassDiff}). This prediction is in conflict with some regional studies \cite<e.g.,>{monsalve2008,bai2013} although in agreement with other global studies \cite<e.g.,>{hacker2015} and regional seismic models \cite{agius2017,gilligan2018}. Additionally, \citeA{pasyanos2014} perturbed CRUST1.0 parameters up to 5\% to fit observed surface waves to make LITHO1.0, which a comparison of CRUST1.0 and LITHO1.0 $V_P$ in the Himilayan region shows that this perturbation reached saturation (i.e. 5\% change between LITHO1.0 and CRUST1.0). If the model was not limited to 5\% change the output would be more felsic than current. Because of the saturation of the method used by \citeA{pasyanos2014}, LITHO1.0 in its current state may not be in agreement with observed regional data and should be revisited. Similar phenomena are observed under the Andes Mountains in South America \cite<e.g.,>{lucassen2001}. The Himilayas are particularly relevant for the geoneutrino prediction signal at Jinping and to a lesser extent, JUNO, while the Andes will be relevant to the future ANDES underground laboratory \cite{bertou2012}. 

\subsection{Geochemical Method for Middle and Lower Crust}
The geochemical method of conversion of $V_P$ to [U,Th,K] does not work well in the lower crust (and to some degree in the middle crust) as the model often surpasses the endmember condition. This is a problem due to simplifying assumptions (i.e. assuming endmembers) indicating a need for a more sophisticated modeling space. A bivariate probability analysis of the available amphibolite or granulite samples with measured $V_P$ could avoid this problem (see Supporting Information for details). This analysis would not assume any specific relationship between $V_P$ and $SiO_2$ (unlike the linear relationship we assumed in this study) but instead would use the bivariate probability of $V_P$ and $SiO_2$ from the dataset. Currently there are too few samples ($\sim$100--150) with measured $V_P$ to create a robust analysis. Incorporation of a thermodynamic modeling software \cite<such as Perple\_X;>{connolly2005}) would allow for the calculation of seismic wavespeed for $>$500 samples, which would significantly increase the robustness of a bivariate probability analysis. Calculated $V_P$ using Perple\_X for samples used in \citeA{huang2013} by \citeA{hacker2015} closely resemble laboratory measurements, indicating the viability of thermodynamically calculated $V_P$. 

\subsection{Near-field (regional) Modeling}
Underestimation of the upper crust uncertainty or problems associated with middle/lower crust abundance calculations have less of an effect on predicted signals if a high-resolution regional geoneutrino model is combined with the global model \cite<e.g.,>{enomoto2007,coltorti2011,huang2014,strati2017}. Because of the distance dependence of the geoneutrino signal (see eq. \ref{eq:flux}) the regional area provides $\sim$40\% of the geoneutrino signal at a detector as calculated in this study. High-resolution seismic and geochemical studies of the near-field therefore provide the most robust estimate of the signal at any detector location. This is exemplified in the studies of \citeA{huang2014} and \citeA{strati2017}, who calculated larger uncertainties on the geoneutrino signal at SNO+ when they included a regional model compared to only using a global model. Although the uncertainty is larger than that estimated from the global model, their estimate includes more local information and is therefore likely to be a more accurate estimate of uncertainty. 

Finally, evolution of our understanding of the physical and chemical nature of the bulk-crust is largely plateauing. This study showed that geoneutrino estimates using LITHO1.0 (2015), CRUST1.0 (2013), and CRUST2.0 (2001) yielded largely the same geoneutrino signal. Furthermore, the geochemical description of the crust from this study is consistent with the estimates by \citeA{rudnick2014}, among others, which is itself consistent with \citeA{rudnick2003}. Because of the natural variability of [U, Th, K] in the crust the uncertainty on bulk abundances is unlikely to change significantly. Joint inversions of multiple datasets, including teleseismic $V_P$ and $V_S$, surface waves, gravity, surface heat flow, and topography/geoid have the most potential for significant advances in our understanding of the structure and composition of the crust (e.g., \citeA{afonso2016,afonso2019}). Regardless, improving our understanding of the physical and chemical nature of the bulk-crust will not have the same impact on reduction in the geoneutrino signal uncertainty as improvements in the near-field. As highlighted by \citeA{strati2017}, better understanding of the physical structure with depth is the most difficult hurdle to overcome in near-field modeling.

\section{Author Contributions}
S.A.W. modeled the geoneutrino signal and wrote the text, with significant input on model creation, data analysis, and the text from O.{\v S}. and W.F.M.

\acknowledgments
Model results and original MATLAB code is provided in the Supporting Material. Support for this study was provided by the University of Maryland Graduate School ALL-S.T.A.R. Fellowship and NSF EAPSI program Award \#1713230 (to S.A.W), NSF grant EAR1650365 (to W.F.M.), and the Czech Science Foundation grant GA\v{C}R 17-01464S (to O.{\v S}.). There are no financial or interest conflicts with this work. We thank Fabio Mantovani, Bed{\v r}ich Roskovec, and Steve Dye for insightful comments and discussion.


%
\bibliography{RMbib}

\end{document}